\documentclass[]{scrartcl}

\usepackage[backend=biber,sorting=none]{biblatex}
\usepackage[affil-it]{authblk}
\usepackage{graphicx}
\usepackage{subfigure}
\usepackage{amsmath}
\usepackage{amssymb}
\usepackage{float}
\usepackage{braket}
\usepackage{lineno}
\usepackage{geometry}
\usepackage{microtype}
\usepackage{siunitx}
\DeclareSIUnit\MIP{MIP}
\DeclareSIUnit\Pixel{Pixel}
\DeclareSIUnit\ADC{ADC}
\usepackage[T1]{fontenc}
\usepackage{lmodern}
\usepackage[noabbrev, capitalise]{cleveref}

\title{Analysis of Testbeam Data Recorded with the Large CALICE AHCAL Technological Prototype}
\subtitle{Talk presented at the International Workshop on Future Linear Colliders (LCWS2021), 15-18 March 2021. C21-03-15.1.}
\author[1]{Lorenz Emberger \thanks{On behalf of the CALICE collaboration}}

\affil[1]{Max-Planck-Institut für Physik}
\affil[ ]{\textit{emberger@mpp.mpg.de}}

\begin{document}

\maketitle

\begin{abstract}
The Analog Hadron Calorimeter (AHCAL) concept developed by the CALICE collaboration is a highly granular sampling calorimeter with \SI{3x3}{\square\centi\meter} plastic scintillator tiles individually read out by silicon photomultipliers (SiPMs) as active material. We have built a large scalable engineering prototype with 38 layers in a steel absorber structure with a thickness of ~4 interaction length. The prototype was exposed to electron, muon and hadron beams at the DESY and CERN testbeam facilities in 2018. The high granularity of the detector allows detailed studies of shower shapes and shower separation with the PandoraPFA particle flow algorithm as well as studies of hit times. The large amount of information is also an ideal place for the application of machine learning algorithms.

This article provides an overview of the ongoing analyses.
\end{abstract}

\section{Introduction}
To fully expolit the potential of future $\mathrm{e^+ e^-}$ collider experiments, a precise reconstruction of all final states is necessary. To achieve this for hadronic final states, a jet energy resolution significantly beyond the current state-of-the-art, in the order of 3\% - 4\% over a wide energy range, is required. Such a resolution, which would provide the capability to separate hadronically decaying $\mathrm{W^{\pm}}$ and Z bosons, requires specialised detectors and reconstruction algorithms. In contrast to traditional energy measurement of summing up the energy depositions in all sub-detectors, the Particle Flow (PF) \cite{Sefkow:2015hna} approach is designed to measure different types of particles of a jet in the best suited detector. For charged particles, the tracker typically offers the best energy resolution. Photon energies are measured in the electromagnetic calorimeter (ECAL) and the energy of neutral hadrons is measured in the hadronic calorimeter (HCAL). Since charged particles and photons carry about 90\% of the energy in a typical jet, particle flow is intrinsically increasing the jet energy resolution. Only the remaining 10\% of the total jet energy carried by neutral hadrons are measured in the HCAL, which typically offers a worse energy resolution. To reduce double counting of charged particles in the tracker and calorimeter, it is vital to separate particle showers in the calorimeters and assign them to tracks, giving rise to a high spatial granularity. This article introduces the beam test data sets taken with the large technological prototype of the CALICE Analog Hadronic Calorimeter (AHCAL) \cite{Sefkow_2019}, a detector designed to complement the particle flow paradigm, and reports on the ongoing analyses. 

\section{Data taking at the SPS}
During two dedicated beam test periods in May and June 2018 the prototype recorded minimum ionizing particle (MIP) tracks of muons as well as particle showers from electrons and pions with beam energies ranging from \SI{10}{\giga\electronvolt} to \SI{200}{\giga\electronvolt}. Additionally, the calorimeter was moved relative to the beam to evaluate the performance over the full front face area and volume. Per particle type and energy several $\mathrm{10^6}$ events were recorded and the detector was operated in both continuous acquisition and power-pulsing\footnote{The power-pulsing mode is designed for the application in a linear collider where the particles arrive in pulsed bunch trains. In the time between two bunch trains the acquisition of the calorimeter is switched off to save power and reduce the dissipated heat. } mode. The MIP tracks characterized by their well defined energy deposition throughout the calorimeter are used for calibration and analysis of the timing performance (\cref{sec:time}). The shower data is used to evaluate the response to hadronically and electromagnetically interacting particles (\cref{sec:shape}) and to study the particle identification performance of the system (\cref{sec:pid}). By artificially superimposing two or more particle showers the separation power, vital for applications in particle flow calorimetry, is studied (\cref{sec:pflow}). All analyses, except the timing performance, are presented using simulated data sets of the full prototype since the analyses are work in progress.

\section{Particle Identification}\label{sec:pid}
\begin{figure}
	\centering
	\subfigure{\includegraphics[width=6cm]{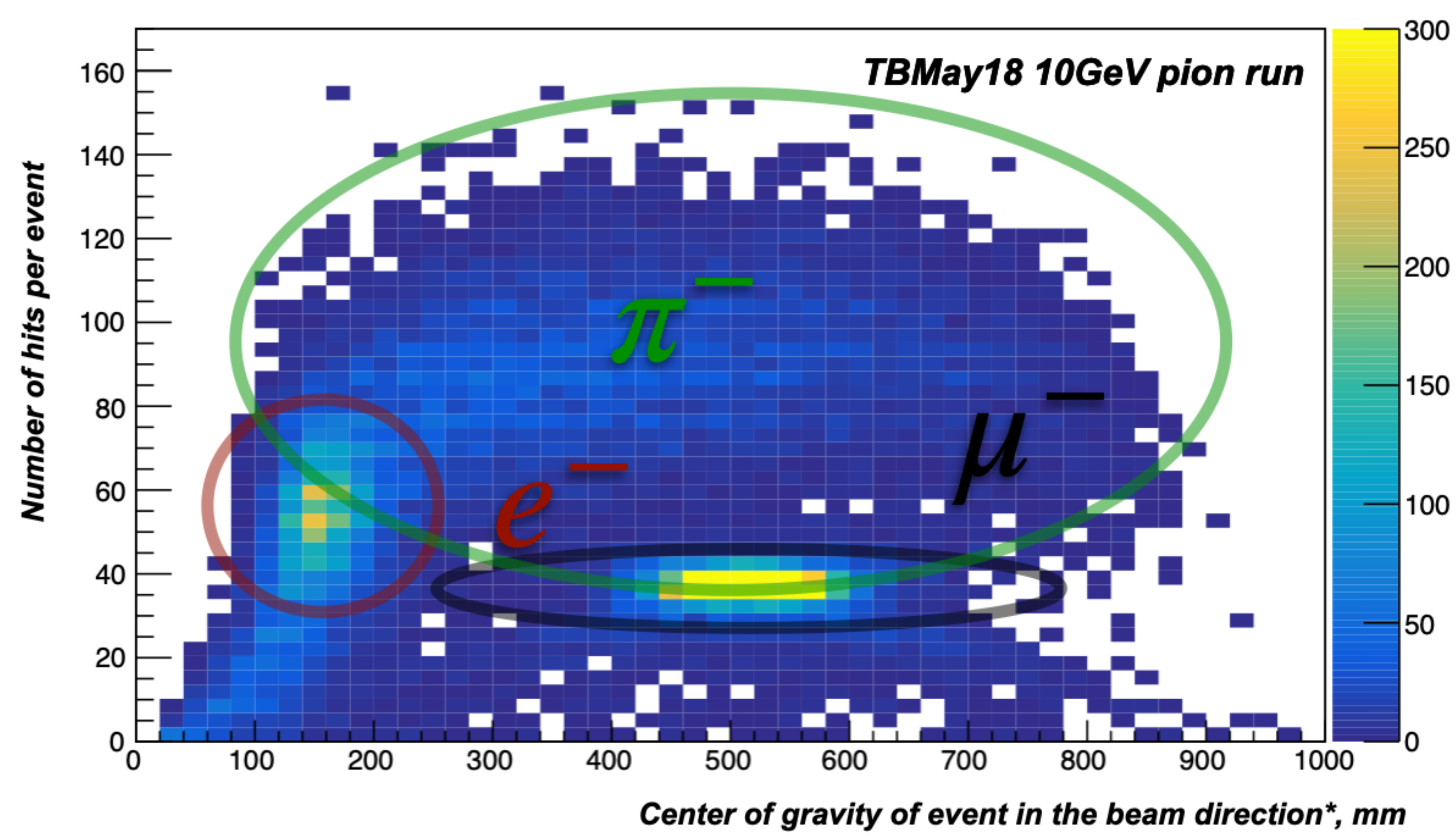}}
	\hspace{1.5 cm}
	\subfigure{\includegraphics[width=6cm]{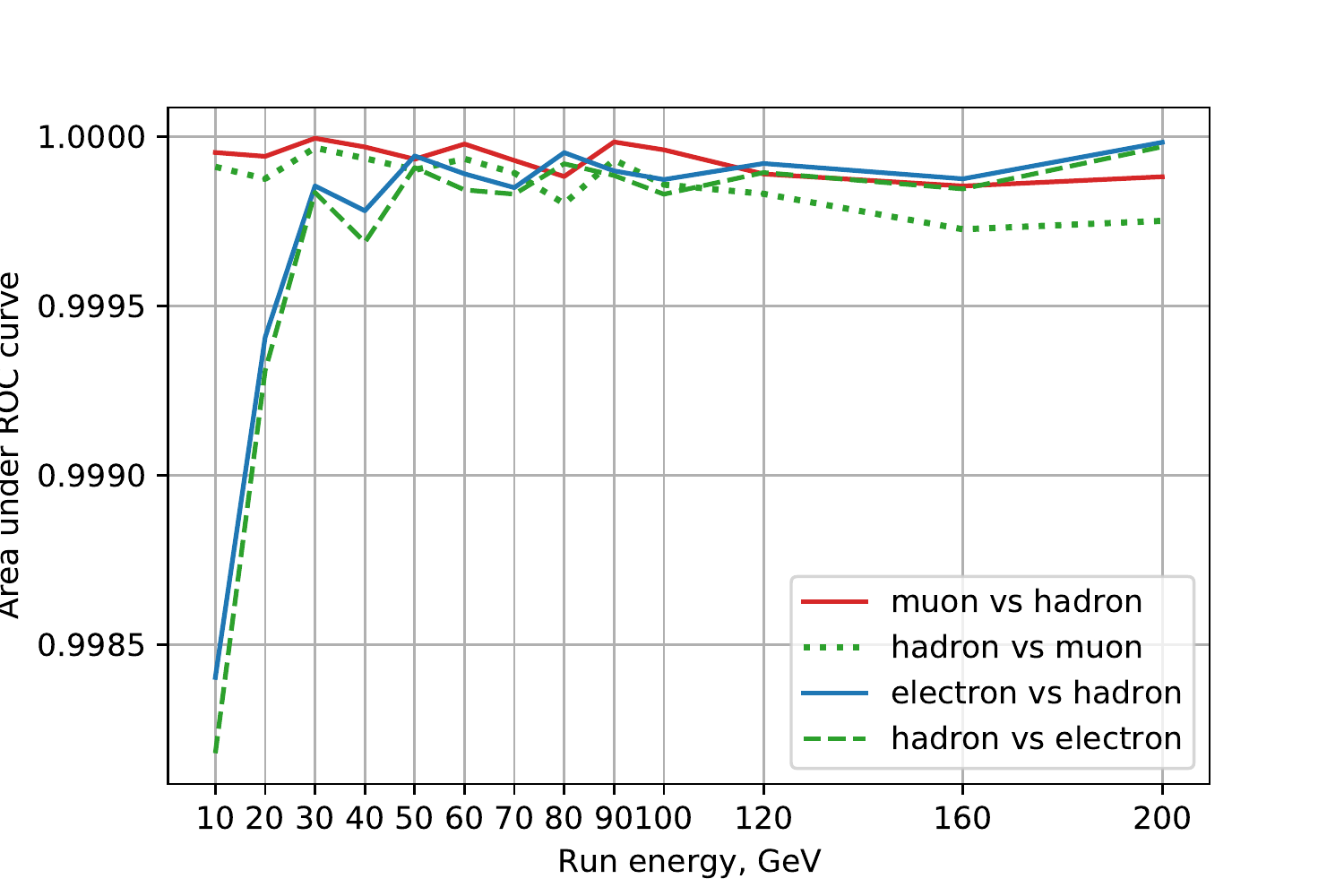}}
	\caption{Left: Center of gravity of the deposited energy in beam direction versus the recorded number of hits per event. The three accumulations are caused by three particle types recorded in the beam test campaign. Right: Particle identification performance of the three classifiers (electron, hadron and muon) in terms of are under the ROC curve for simulated data. An excellent separation is achieved over the full energy range \cite{PrivCommBoch}.}\label{nhvsz}
\end{figure}

Efficient identification of particles is a key ingredient to successful particle flow algorithms to reduce the danger of confusion and double counting of particles. The left image in \cref{nhvsz} shows the distribution of the center of gravity along the beam direction versus the number of hits in the calorimeter for a mixture of particles. As indicated in the picture, three major accumulations are visible corresponding to three types of particles. 
Since these accumulations overlap, additional observables have to be used to achieve sufficient particle identification giving rise to a multivariate approach. A boosted decision tree (BDT) is implemented for every particle type and trained on labeled events obtained with a simulation of the full prototype. The chosen features contain information on the center of gravity, the longitudinal and transversal development of the particle showers, the longitudinal position of the shower start, the fraction of deposited energy contained in showers and tracks and the number of hits contained in showers and tracks. Each BDT is trained on a simulated sample containing only its respective particle type and tested on a simulated data set containing a mixture of all three particle types. The right image in \cref{nhvsz} parameterizes the separation power of the classifiers over the simulated energy range, omitting the case of electron versus muon because the potential for confusion is larger in the shown cases. The measure of separation power is the area under the receiver operating characteristic (ROC)\footnote{A ROC curve describes the performance of a classification model at all classification thresholds by showing the ratio of false positive rate to true positive rate. The integral of this curve measures the identification power of the model, where 0 characterizes a 100\% wrong model and 1 characterizes a 100\% correct model.} curve showing excellent performance over the studied energy range. 

\section{Shower Shape Analysis}\label{sec:shape}
\begin{figure}
	\centering
	\subfigure{\includegraphics[width=6.3cm]{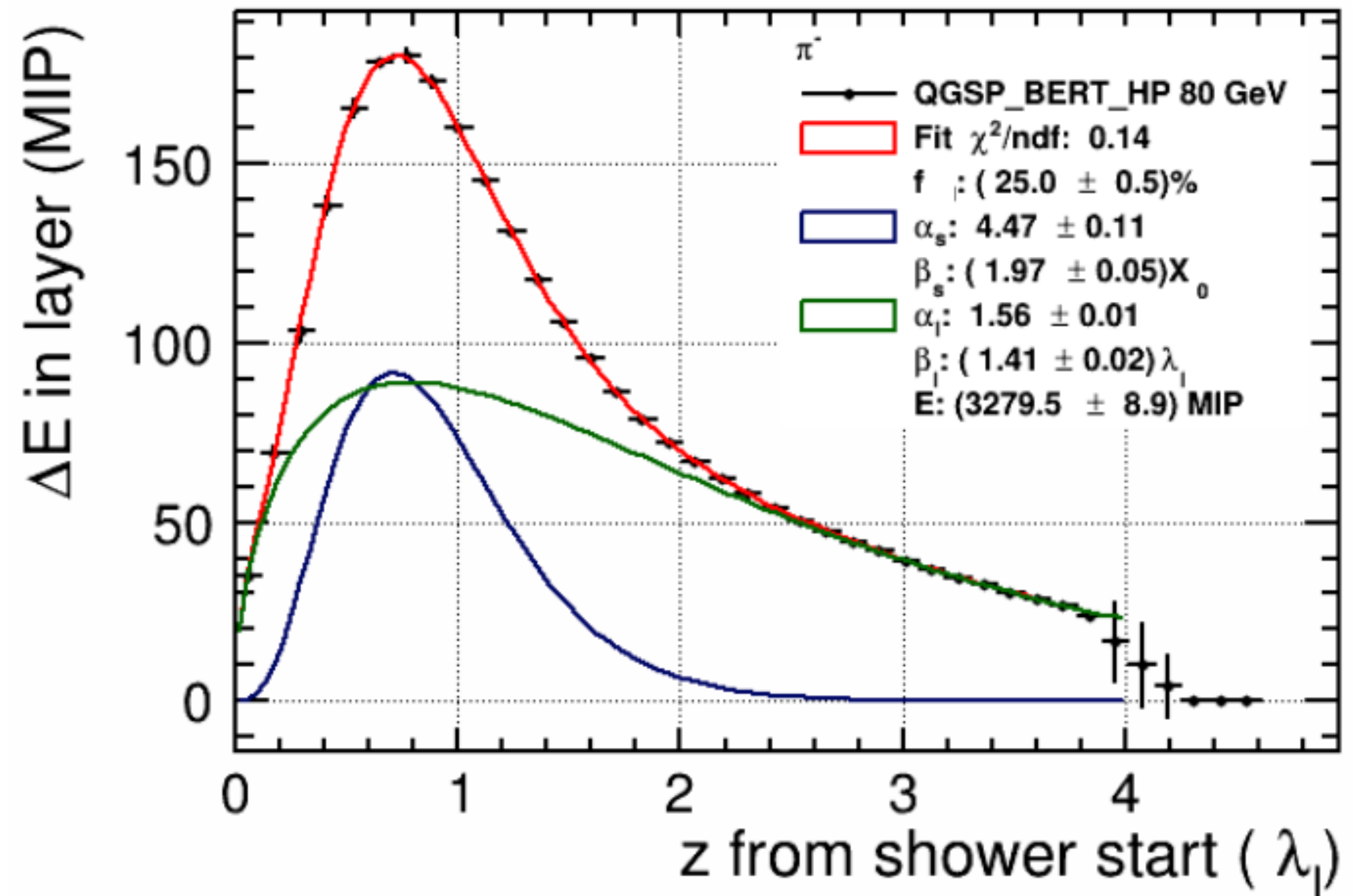}}
	\hspace{1.5 cm}
	\subfigure{\includegraphics[width=6.3cm]{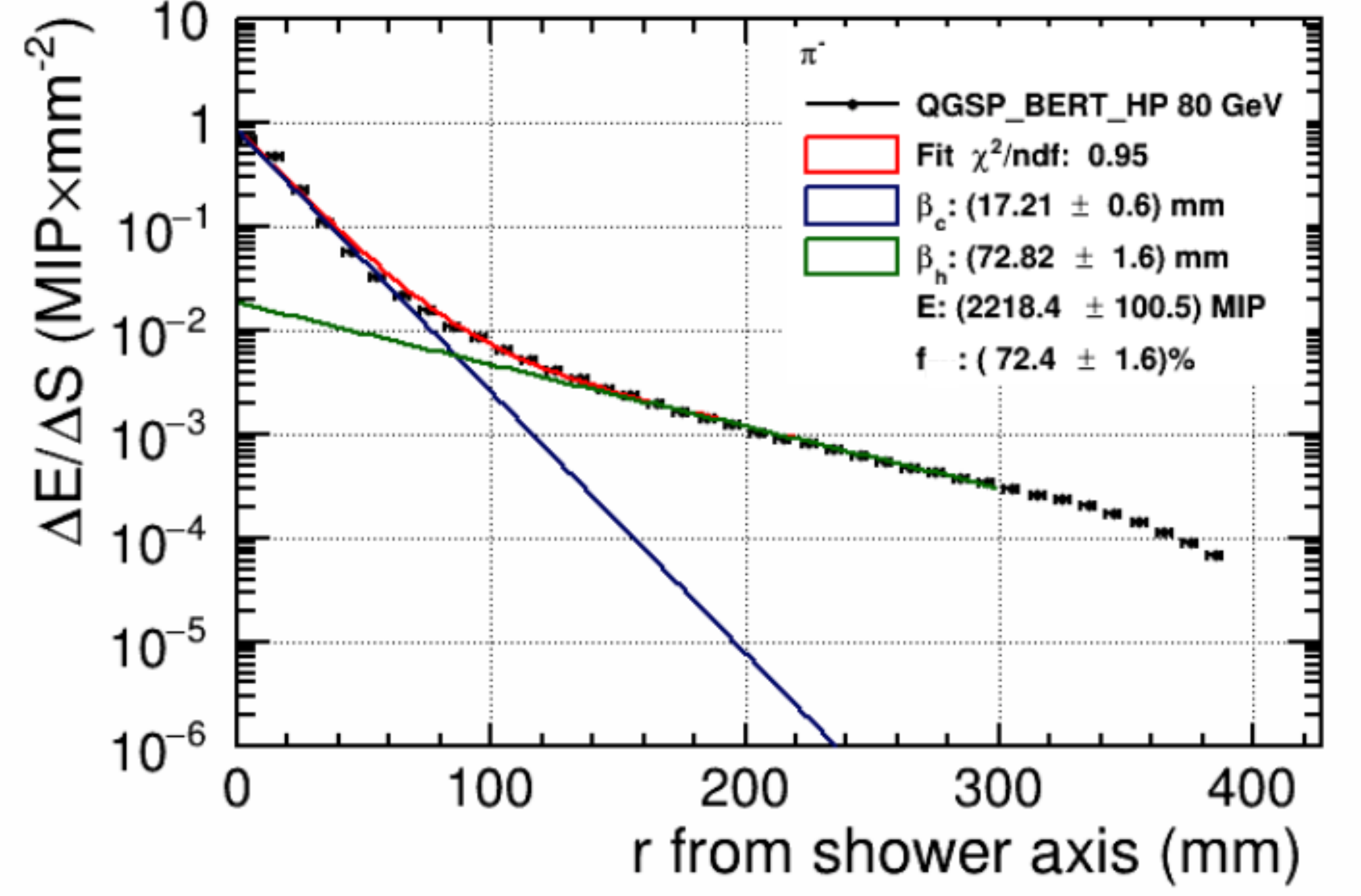}}
	\caption{Left: Longitudinal shower development of simulated \SI{80}{\giga\electronvolt} Pions with fitted parameterization (1). Right: Radial development of simulated \SI{80}{\giga\electronvolt} Pions with fitted parameterization (2) \cite{PrivCommPinto}.}\label{shapes}
\end{figure}

The high spatial granularity and longitudinal segmentation of this prototype offers a unique possibility to study the shape of hadronic showers. 
By investigating the longitudinal and radial energy density, the shower can be separated into the electromagnetic core and the halo part. The longitudinal parameterization is based on the assumption that the evolution of the core part is governed by the radiation length $\mathrm{X_0}$ and the halo is governed by the nuclear interaction length $\mathrm{\lambda_l}$. It is composed of a sum of incomplete gamma functions and reads
\begin{equation}
	\Delta E(Z)=E\cdot\left( \frac{f}{\Gamma(\alpha_s)}\cdot\left( \frac{Z[X_0]}{\beta_s}\right)^{\alpha_s -1}  \cdot \frac{e^{\frac{Z[X0]}{\beta_s}}}{\beta_s}  + \frac{1-f}{\Gamma(\alpha_l)} \left(   \frac{Z[\lambda_l]}{\beta_l} \right)^{\alpha_l - 1} \cdot \frac{e^{\frac{-Z[\lambda_l]}{\beta_l}}}{\beta_l} \right) .
\end{equation} 
Here $\mathrm{\alpha_s}$ and $\mathrm{\beta_s}$ are shape parameters of the core component, $\mathrm{\alpha_l}$ and $\mathrm{\beta_l}$ are the shape parameters of the halo component, $\mathrm{f}$ is the relative weight of core to halo component, $\mathrm{Z[\lambda_l]}$ is the depth in the calorimeter in terms of the nuclear interaction length $\mathrm{\lambda_l}$ and $\mathrm{Z[X_0]}$ is the depth in the calorimeter in terms of the radiation length $\mathrm{X_0}$. The radial profile is parameterized by an exponentially decreasing energy density in concentric rings of area $\mathrm{\Delta S = 2\pi r \Delta r}$ and reads

\begin{equation}
	\frac{\Delta E}{\Delta S}\left(r\right) = \frac{E}{2\pi}\left( f\cdot \frac{e^{\frac{-r}{\beta_c}}}{\beta^2_c} + (1-f) \cdot\frac{e^{\frac{-r}{\beta_h}}}{\beta^2_h} \right),
\end{equation} 
where r is the radius from the shower center and $\mathrm{\beta_c}$ and $\mathrm{\beta_h}$ are the shape parameters of core and halo part. The simulated shower shapes including the fitted parameterizations are shown in \cref{shapes}.
From the weight of core and halo, f, the fraction of energy deposited in the respective part of the shower can be calculated from the total deposited energy. This study was also performed for the previous AHCAL physics prototype \cite{Eigen_2016} and is currently repeated for the large technological prototype to exploit its higher spatial granularity and the reduced noise.

\section{Particle Flow}\label{sec:pflow}
\begin{figure}
	\centering
	\subfigure{\includegraphics[width=6cm]{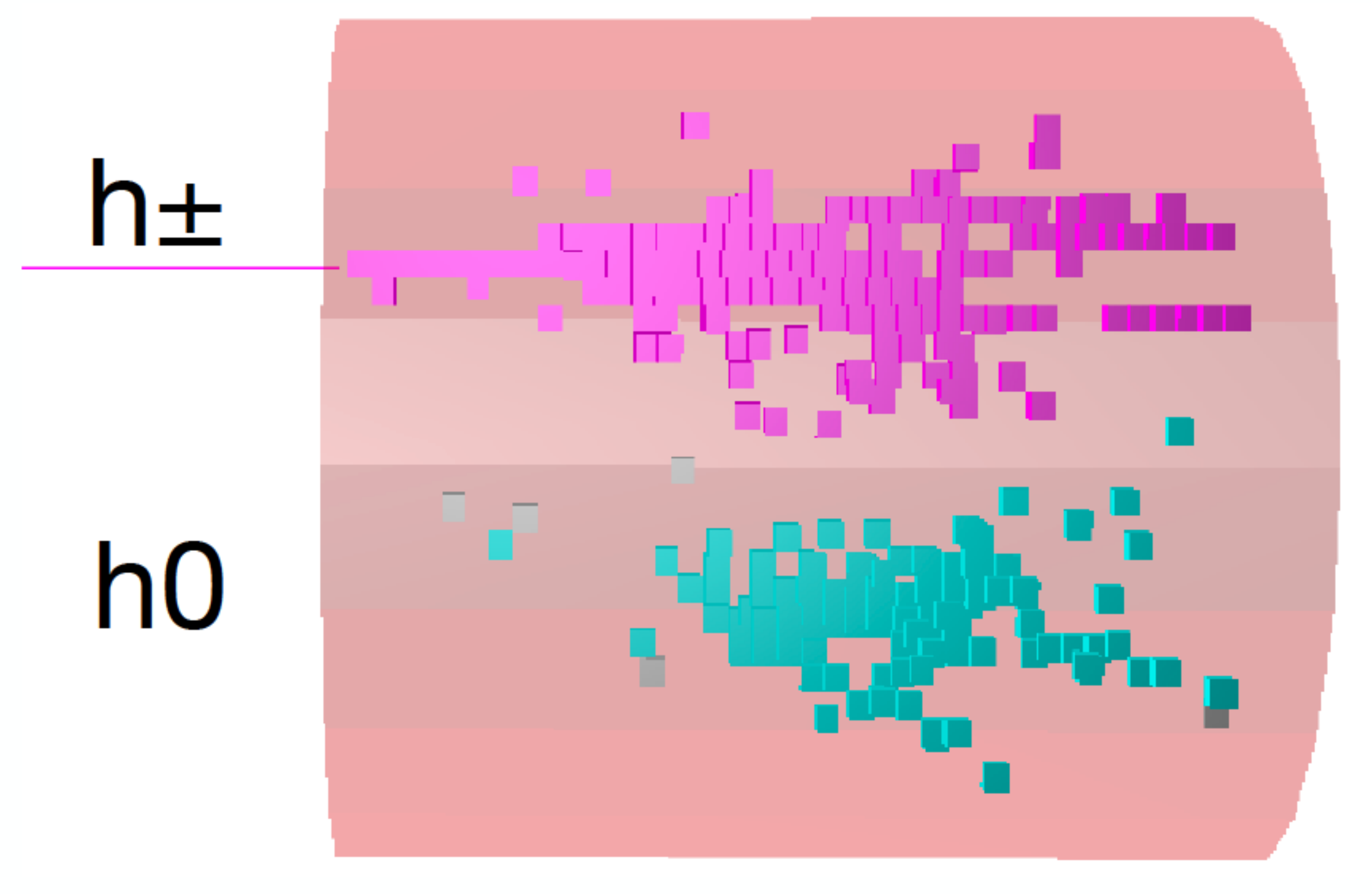}}
	\hspace{1.5 cm}
	\subfigure{\includegraphics[width=6.3cm]{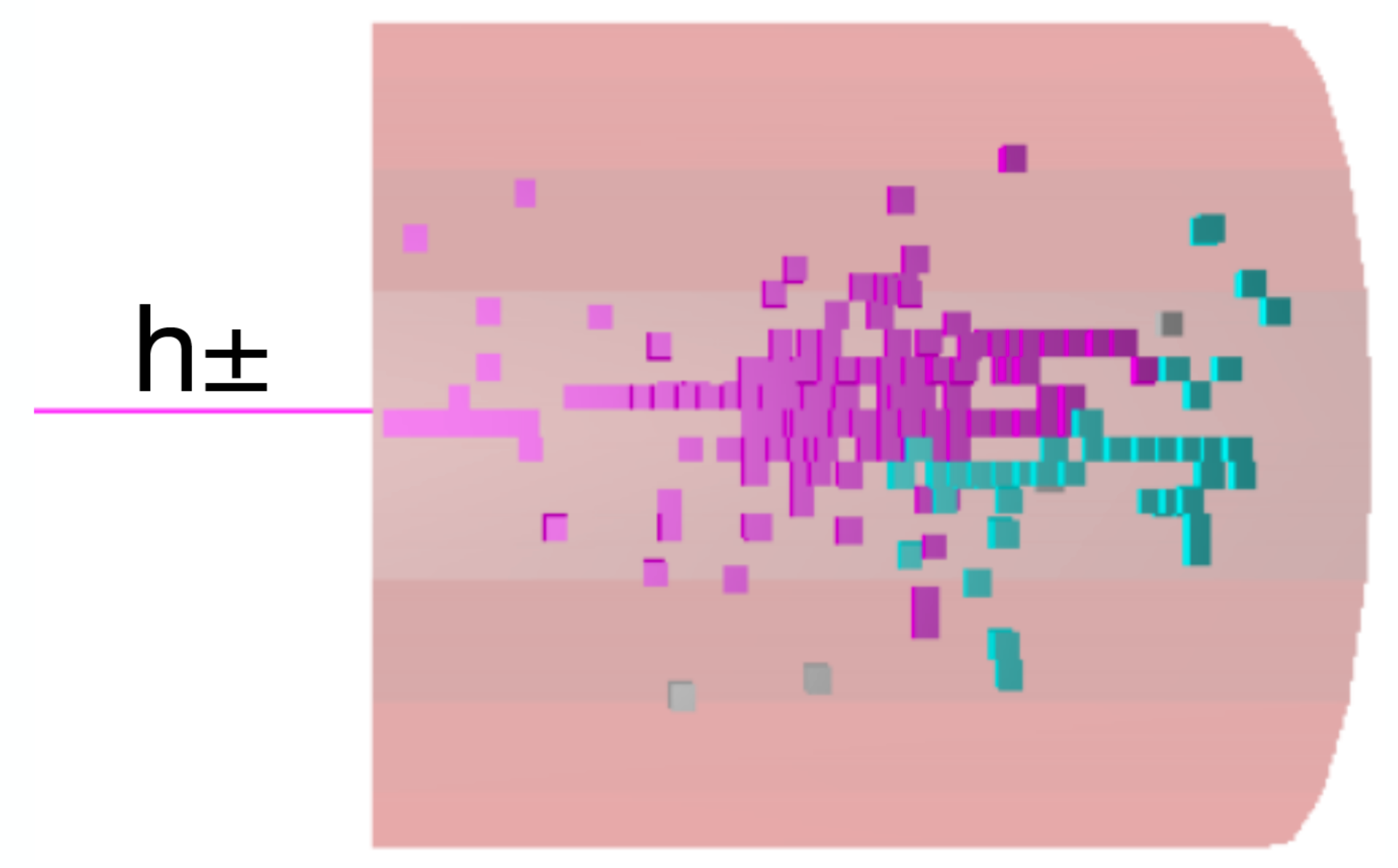}}
	\caption{Left: Overlayed event containing a charged hadron (h$\mathrm{\pm}$) and a artificial neutral hadron (h0) generated by removing the primary track. Pandora correctly identified the neutral shower(cyan) and the charged one(magenta). Right: Partial misclassification of a single particle event. The algorithm produced an excess of neutral energy(cyan) although only one charged hadron entered the calorimeter \cite{PrivCommHeuchel}.}\label{pflow}
\end{figure}

The main task of a hadronic calorimeter in the particle flow paradigm is the reconstruction of neutral hadrons within particle showers. In reality, charged hadrons will also reach the calorimeter so a reconstruction algorithm has to be implemented, which is capable of separating the showers and associating them to their respective mother particle. To reproduce and study this scenario with the AHCAL, two hadron events have to be overlayed. Since neutral hadrons don't leave a primary track in the calorimeter before the shower starts to develop, this track has to be removed from the event to generate a fake neutral hadron. The overlayed events are presented to the full Pandora particle flow algorithm chain \cite{Thomson2009pandora} to identify the two showers. An example of the resulting events is shown in \cref{pflow} on the left.  
The algorithm is able to do a correct separation for the majority of the studied events, also considering variations in the relative particles energies and proximity. The right image in \cref{pflow} shows a rare case of misclassification in which only a single particle was presented to Pandora, but the resulting event shows an excess of neutral energy. Events of this kind are currently under investigation. This study was previously performed using data from the AHCAL physics prototype \cite{Adloff:2011ha} and is currently repeated to exploit the higher spatial granularity and reduced noise of the large technological prototype.

\section{Analysis of the Timing Performance for MIPs}\label{sec:time}

\begin{figure}
	\centering
	\subfigure{\includegraphics[width=6cm]{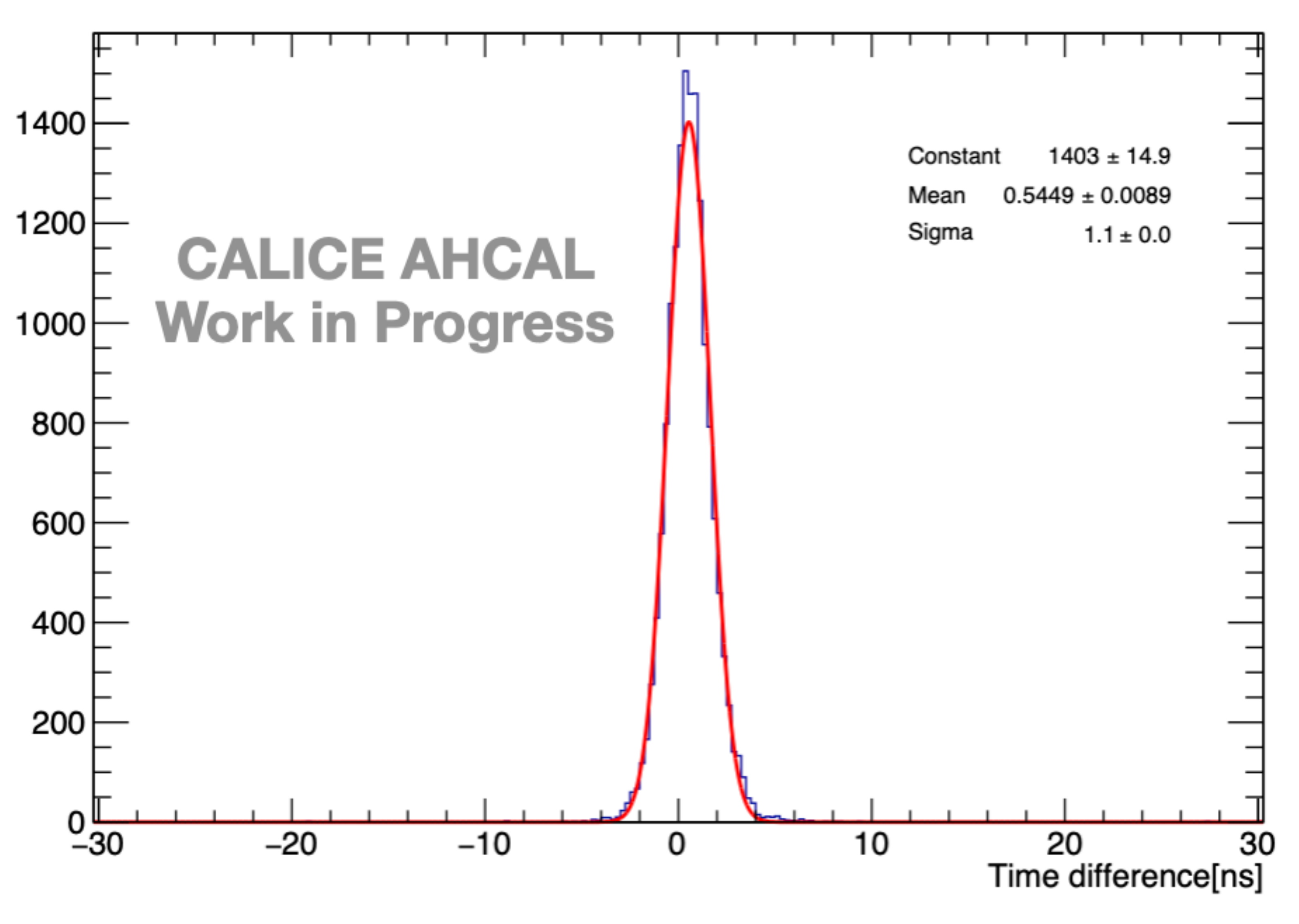}}
	\hspace{1.5 cm}
	\subfigure{\includegraphics[width=6.4cm]{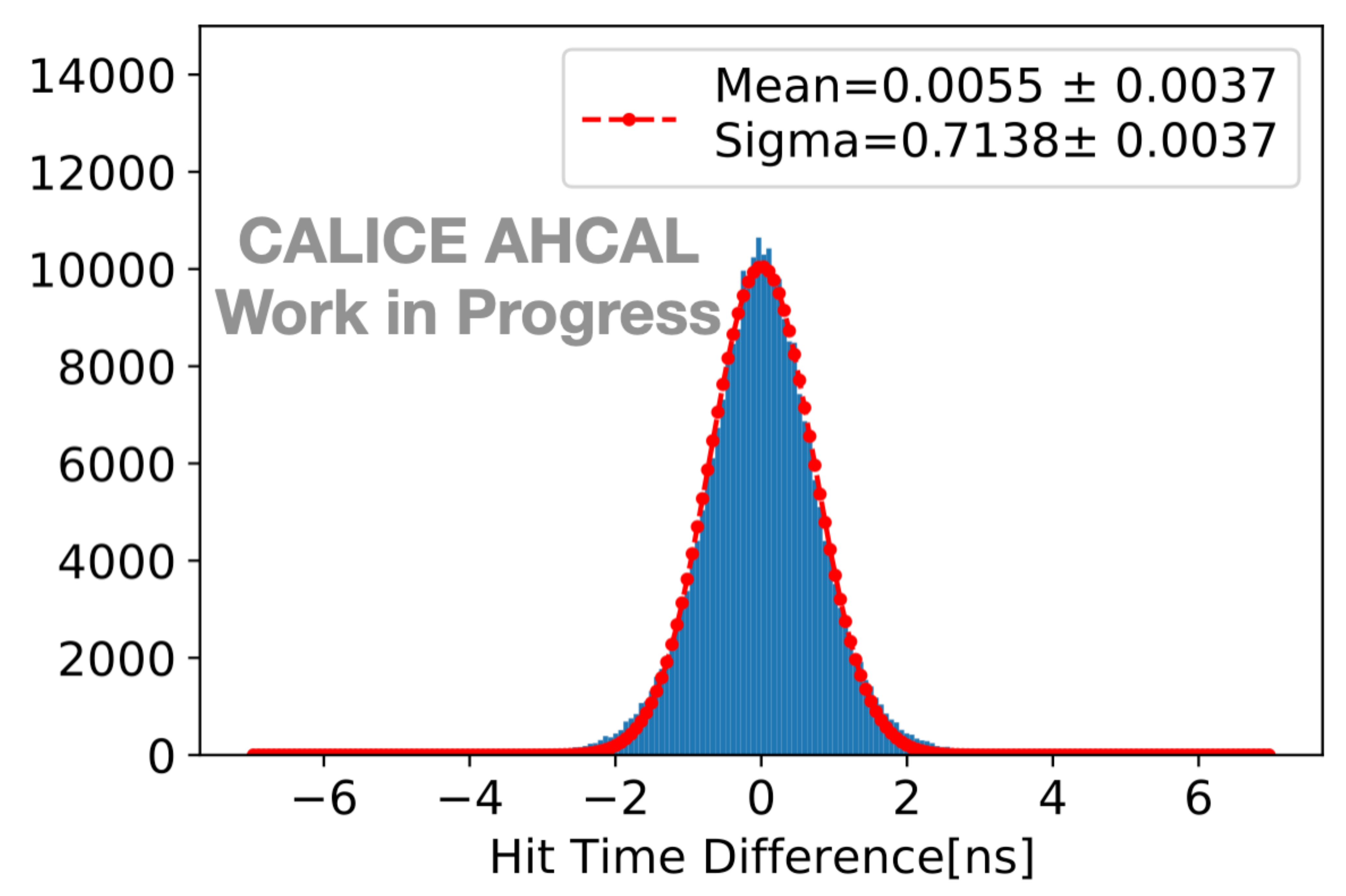}}
	\caption{Left: Distribution of the hit time difference in two subsequent channels of the AHCAL. Right: Distribution of the hit time difference obtained with the dedicated SiPM-on-Tile timing setup \cite{PrivCommLenz}.}\label{time}
\end{figure}

In addition to the hit energy measurement, the AHCAL is also capable of performing single channel hit time measurement enabling the separation of particles on the basis of their arrival time at the calorimeter. Furthermore, since the time of the interaction in a particle collider is well defined, this information can be used to reject out-of-time background providing clean events for high precision physics. The design time resolution of the AHCAL is \SI{1}{\nano\second} or below for minimum ionizing particle (MIP) tracks penetrating the calorimeter and causing one hit in every layer. The data for this analysis was recorded at the DESY test beam facility \cite{DIENER2019265} in 2019. The timing performance is investigated by calibrating the hit time against an external trigger time and obtaining the time difference of two subsequent channels. The resulting distribution is shown in \cref{time} on the left. In order to obtain the single channel time resolution from this distribution, its width has to be divided by $\mathrm{\sqrt{2}}$ resulting in a resolution of \SI{780}{\pico\second}. Since the time measurement in the AHCAL is affected by the involved electronics and calibration procedure, a dedicated beam test setup was designed to measure the intrinsic time resolution of the SiPM-on-Tile configuration as it is used in the prototype. It consists of four individual channels using the same scintillating tiles and SiPMs but with simplified electronics and a fast digitizer to record the full analog waveform at a sampling rate of \SI{2.5}{\giga\hertz}. The tiles are arranged in a beam telescope-like setup, the outer two tiles serve as coincidence triggers while the inner tiles are used to perform the time measurement. This setup was tested at DESY in 2020. Similar to the analysis performed for the AHCAL, the hit times of the inner tiles are obtained relative to the coincidence trigger time and subtracted from each other. The resulting distribution is shown in \cref{time} on the right, the intrinsic time resolution of the SiPM-on-Tile setup was determined to be \SI{507}{\pico\second}. This value implies that the front-end electronics of the AHCAL contribute about $\mathrm{\sqrt{780^2ps - 507^2ps}=}$\SI{593}{\pico\second} to its single channel time resolution\footnote{The measurements leading to these results have been performed at the Test Beam Facility at DESY Hamburg (Germany), a member of the Helmholtz Association (HGF).}.

\section{Summary}
This article summarizes the currently ongoing studies performed on the data taken with the large CALICE AHCAL technological prototype in 2018. It shows the excellent particle identification and separation power using multivariate classifiers and the Pandora particle flow algorithm. Studies investigating the calorimeters response to the components of hadronic showers are ongoing and show promising results on simulated data. Dedicated data sets taken at DESY in 2019 and 2020 are used to investigate the timing performance of the prototype showing that the design goal of \SI{1}{\nano\second} for minimum ionizing particles was reached. 
\section{Acknowledgment}
I would like to thank the AHCAL group within the CALICE collaboration, in particular Vladimir Bocharnikov, Olin Lyod Pinto and Daniel Heuchel for contributing their research to this article. 
\printbibliography

\end{document}